\documentstyle[aps,epsfig,multicol]{revtex}
\begin{document}
\draft
\title {Quantum fluctuation induced spatial stochastic resonance  at zero temperature }
\author{Parongama Sen}

\address{Department of Physics, University of Calcutta,
92 A.P. C. Road, Calcutta 700009, India.\\
e-mail paro@cucc.ernet.in}
\maketitle


\begin{abstract}

We consider a model in which the quantum fluctuation can be controlled and 
show that the system 
responds to a spatially periodic external field at  zero temperature. 
This signifies  the occurrence of spatial stochastic resonance where the
fluctuations are purely quantum in nature.
 Various features of the phenomenon are discussed.\\
\end{abstract}
\begin{multicols}{2}
\medskip

\medskip

The phenomenon of stochastic resonance (SR) 
is precisely the enhancement of reponse to an external 
field with the help of noise. SR is manifested in bistable systems, where the
fluctuations can drive the system from one energy minimum to the other.
As a function of the noise, the response typically shows a maximum value 
at resonance. The basic features required to observe stochastic resonance
are simple: an energy barrier, a weak external periodic signal and a noise 
source. 
SR, which is observed in  many natural phenomena,  has therefore important 
applications  
in a variety of research areas including non-linear optics, solid state devices,
and even  neurophysiology \cite{rmp}. 

Although most of the studies  consider classical systems,
stochastic resonance in   quantum systems have attracted a lot of 
attention recently \cite{rmp,gross,LC,Grif}. 
In the quantum systems, the quantum mechanical 
tunnelling provides an additional channel to overcome a potential barrier.
At exactly zero temprature, some numerical results for the tunnelling
phenomena are available for the 
periodically driven quantum 
double well system  \cite{gross}. 
The noise is purely quantum in nature at $T = 0$. A systematic
study of the variation of a response function with the noise
is not available here due to the inherent complexities of the dynamical system
\cite{gross}; e.g., tunnelling is enhanced for high and 
low frequency limits but  it may altogether be destroyed (coherent destruction
of tunnelling) in the intermediate range. 
At very low temperatures, where quantum fluctuations still dominate,
investigation on quantum stochastic 
resonance (QSR) is effectively reduced to the study of the 
dissipative dynamics  of a periodically driven spin-boson system 
\cite{rmp,Grif}. 
 The noise here is characterised by the temperature of the thermal bath 
and by the coupling of the bistable system to the environment. 

Recent results in some classical systems have shown that the 
the positive role of  noise to enable  response to  a signal can  
by no means be restricted to temporal fields only
and is more universal in nature  \cite{gade,vilar}.
Even when the field is spatially periodic, it 
can induce  spatial modulations in a bistable system  with the help of noise. 

We report  here that one can  achieve QSR at $T=0$ 
in a  novel quantum system, in which the response  can
be studied as a function of the quantum noise. 
Motivated by the occurrence of spatial SR  in classical systems,
we take the external field to be periodic in space in the quantum case.
Our system is  a periodically driven Ising model in a transverse field at 
zero temperature.  
The strength of the transverse field is a measure of
the quantum fluctuations and can be tuned.

Although it occurs in both cases, the mechanisms of SR
in  fields periodic in time  and spatially modulated  
fields  are intrinsically different.
In   stochastic resonance in presence of time dependent 
fields, there are resonant transitions between the two potential
wells of the bistable system. It is acheived when the time period of 
the temporal field
is  comparable to twice the systems's own time scale of 
transitions between the neighbouring potential wells, the rate of which  
is given by the Kramer's rate. 
In presence of a spatially periodic field, on the other hand,
resonance implies  transitions to a state with 
 spatial correlations  commensurate to those of the 
external field. The  symmetric double well potential  
gets distorted  differently  in the two cases, although fluctuation induced 
transitions are responsible for  resonance in both.  
In the spatially periodic field,  the system reaches an equilibrium and studying
the static properties are sufficient.
Instead of time scales as in time dependent fields, one has to compare the
length scales in the spatially modulated field.
The  field   tends to 
form domains of size of the order of half of its wavelength. 
At resonance, the correlation length of the equilibrium state
has to be comparable to  half 
the spatial periodicity or wavelength of the field. 

The lattice periodicity plays an important role in a spatially periodic field.
The wavelength of the field may 
be either commensurate or incommensurate to the lattice periodicity. 
In the present study, we consider the commensurate case only.
The case of incommensurate field has not been studied for 
the classical case even.

The  transverse Ising system in a spatially periodic field 
can be described by the Hamiltonian
\begin{equation}
H = -J\sum _{i=1}^{N}S^{z}_{i}S^{z}_{i+1} -\sum_{i=1}^{N}h_iS^{z}_{i}
-\Gamma \sum_{i} S^{x}_i.
\end{equation}
Here $h_i$ is the  spatially modulated field and $\Gamma$  the 
transverse field.
The form of $h_i $ is like this: $h_i = h_0$ at $i=(n\lambda +1) $ to 
$(2n+1)\lambda/2$ ($n=0,1,....,N/\lambda -1$) and $-h_0$ elsewhere. 
 The  wavelength of the field is 
 denoted by $\lambda$.
In principle  the modulated field can be chosen 
to be of any   form, for simplicity we  choose a square wave form.
The Hamiltonian (1) is comparable to that of the  transverse Ising 
model in a field periodic in time \cite{miyashita,CDS} 
which is studied in the context of quantum dynamical phenomena. 
We consider a spatial equivalent and 
will be commenting on the comparable features later in this paper.

In the absence of the longitudinal field, the system 
has a quantum critical point at $\Gamma/J =1$ \cite{pfeuty}.
With the field, which is competing in nature, the phase 
transitions in the $\Gamma-h_0$ plane 
have been recently 
studied \cite{sen}. The system shows a continuous order-disorder
phase transition from 
$\Gamma/J = 1$ at $h_0 =0$ to $\Gamma/J =0$ at $h_0 = 4J/\lambda$. Beyond this 
value of $h_0$, the system is field dominated and no phase transition can occur 
here. Thus our study will be  limited to the
region $h_0 < 4\lambda/J$, beyond which 
the  system responds to the field spontaneously.
 $\Gamma =0, h_0 = 4J/\lambda$ is a multiphase point. 
The model has the additional feature  that the competing field and the 
tunnelling field scale similarly close to criticality.

We obtain the  average  correlations between the local fields and spin
components in the longitudinal ($z$) direction: 
\begin{equation}
g(h_0, \Gamma ) = \frac{1}{N}[ \frac {1}{h_0} { \sum_i\langle S_i^z h_i\rangle }], 
\end{equation}
$\langle ... \rangle $ is the expectation value and consider it as the 
response function \cite{gade}. $g(h_0, \Gamma ) \leq 1$ by definition.  
Resonance will imply a maximum value of $g(h_0,\Gamma )$.

Stochastic resonance is a process induced  by fluctuation.
In a system which undergoes a continuous phase transition, the fluctuations
are maximum at the critical point. However, SR is expected  away from  
criticality where fluctuation is lesser. This is because at the 
critical point, the correlation length diverges and all other length 
scales are irrelevant. But for spatial SR to occur,  the spin
correlation length  should equal the imposed spatial
modulation. The latter being finite, SR will occur in a regime away
from the critical point
This is confirmed in our results.


We conduct a numerical study by diagonalising the Hamiltonian matrix
for finite chains using Lanczos method. 
The system sizes are restricted to  multiples of
the   wavelength $\lambda$ of the field. 
We obtain results for  system sizes $N \leq 18$ with  periodic boundary 
conditions.
 In Fig. 1, we show  how the spins orient spatially (the $z$
components of the spins are shown, to which the external
periodic field is coupled) as 
the noise is increased for a constant value of $h_0$. It is interesting to 
note that even when the fluctuations are small so that a ferromagnetic 
order exists, the local magnetisation shows a modulation with wavelength 
$\lambda$ about 
a non-zero value.  Beyond the ferromagnetic phase, it oscillates 
about zero. The modulations 
will increase with the noise, as expected and then decrease beyond a certain
point. 

In fig. 2, we show how the response function  $g(h_0,\Gamma )$ behaves with 
the increasing quantum fluctuation  ($\Gamma /J$): it  shows a 
maximum at a certain value of $\Gamma = \Gamma_R$. 
The position of the maximum and the  value of $g$ depend on the
wavelength as well as on  the field strength.

The behaviour of $\Gamma_R/J $ with the field $h_0/J$ is shown in Fig. 3 
for different values of the wavelength. 
 $\Gamma_R$ is not defined for $h_0 = 0$ and  
$\Gamma_R = 0$ for $h_0/J \geq 4/\lambda$ where the system
 naturally orients along the field. 
It may be noted that as the 
wavelength is increased, the resonance occurs at lower  values of 
$\Gamma$. This is quite easy to understand:  for increasing wavelength 
$\lambda$,
one needs to create smaller number of domain walls and hence lesser
amount of fluctuation is needed. 
The phase boundaries ($\Gamma_c(h_0)$)  show opposite behaviour as 
with increasing $\lambda$, one needs larger amount of fluctuations to 
destroy the order \cite{sen}.
For a  comparison
of $\Gamma_c$ and $\Gamma_R$, we have also shown the phase boundary line
for $\lambda =2$ in Fig. 3.

Another quantity which we study is the value of the parameter 
$g(h_0, \Gamma )$ 
at $\Gamma_R$. This shows a very interesting behaviour.
It varies linearly with $h_0$ upto a limiting  value
and then shows a  nonlinear behaviour, increasing faster than a power
law (Fig. 4). This can be interpreted as the
existence of a limiting value of $h_0$ beyond which the
field can no longer be assumed to be weak. This limit also
decreases with $\lambda$. 
Beyond this limit, the unperturbed energy landscape picture seems to get 
drastically modified.  
It is also interesting to note that even for strong fields, while the maximum 
response is less than 0.6
for $\lambda = 2$ and 4, 
 it rapidly approches  unity for higher values of $\lambda$. 

In order to study how the value of $\Gamma_R/J $ varies with the wavelength, we 
plot $\Gamma_R/J $  against $\lambda$ for the same   value of $h_0\lambda /J $. 
 The reason for keeping $h_0 \lambda /J $ constant is that 
$h_0 \lambda $ determines  the limiting value
of $h_0$ and could be used as a standard to compare results
corresponding to different values of 
$\lambda$. The results are shown in Fig. 5. 
For  values of $h_0\lambda/J $ much lesser than  4,  $\Gamma_R/J $  
shows a monotonic behaviour.
For larger values close to the multiphase point, there
is an anomaly and the monotonic behaviour is lost in the sense 
$\Gamma_R/J$ first increases with $\lambda$ and regains its decreasing nature
beyond $\lambda =4$. Our explanation for the above is that for very low
value of $\lambda$, the field amplitude is very strong at high values of 
$h_0\lambda$ and thus 
one needs a smaller amount of fluctuation to achieve resonance. 
This is again a feature of the strong field regime. 
Perhaps $h_0\lambda /J$ is not the proper parameter
to keep constant and compare results here.
Apparently,  $\Gamma_R/J$ decays expoenentially with 
$\lambda$ for large $\lambda$ values in the weak field regime.

At the multiphase point, the domain sizes are multiples of $\lambda /2$.
Close to the multiphase point,  one could expect correlations other than that of the field
to manifest.  We investigate this 
for high values of $h_0$  close to $4J/\lambda$  but 
fail to detect any other modulations barring that of $\lambda$.
Thus, any
  nontrivial  behaviour of the magnetisation (e.g., step like structures etc.), 
is definitely a
 feature unique to time dependent fields \cite{miyashita,diptiman}, 
as was emphasised in \cite{miyashita}
and absent for spatially periodic fields.

The results shown in  figures 1-5 are for $N=12$ for $\lambda = 12$, 
$N=16$ for $\lambda$ values
2, 4, 8 and 16 and $N=18$ for $\lambda = 6, 18$.
The finite size effects are negligible and not shown.

The general features of the spatial stochastic resonance occurring in the 
quantum Ising model at zero temperature are quilitatively  similar 
to those of the 
classical Ising model \cite{gade}.  
The comparison of $\Gamma_R$ and $\Gamma_c$ is a unique feature of 
the  quantum system, as no finite temperature
phase transition exists in the classical one-dimensional model.
However, stochastic resonance does not require that the system
should undergo a phase transition.
Also, it is not imperative that SR in one dimensional quantum
sustem and two dimensional classical system be identical as SR is not 
a critical phenomena.

In summary, we have studied the phenomenon of spatial stochastic resonance in 
a quantum model at $T = 0$ and found that SR can be realised in this model
where the fluctuations are entirely quantum in nature. The response 
of the system depends on the field amplitude and the wavelength.
 Resonance is achieved  at 
values of the transverse field higher
than its critical value where the order-disorder transition occurs.  
The maximum response behaves differently in the  so-called weak field 
and strong field regimes.  
We could not detect any oscillations of the local magnetisation
with periodicity different from that of the external field  even 
close to the highly degenerate multiphase point.

The reason we could observe  quantum stochastic 
resonance at zero temperature are  twofold: 
firstly, the 
quantum model we consider may be special.
Secondly, the spatial and temporal fields  signify entirely different
processes of stochastic resonance. Whether it is possible to 
obtain QSR in a field periodic in time for the transverse Ising model
will be an interesting study and confirm whether the first
conjecture is correct.  
 The ground states of the transverse Ising model
and the quantum double well system are equivalent and the
counter-intuitive feature like coherent destruction
of tunnelling is present in both \cite{gross,miyashita} when a  field
periodic in time is present.
Also, a  simple picture like  Fig. 2 is unlikely to exist    
in presence of  a time
dependent  periodic field, as there are  
indications of additional nontrivial oscillations of the
magnetisation in quantum spin models \cite{miyashita,diptiman}.
Hence it seems that choosing the field periodic in space, rather
than considering  a specific quantum system, is  responsible for  
observing  quantum stochastic resonance analogous to the classical case.


Although we are interested in the static results only for SR, 
there will be an inherent
quantum dynamics in the model. 
The dynamics of  domain growth  
could  be an interesting study which in this case may be called 
 a local nucleation 
phenomena;  the domains will be  finite in size 
due to the presence of the spatially periodic external field. 
\bigskip

The author is grateful to S. M. Bhattacharjee for a critical reading of the
manuscript and to B. K. Chakrabarti for discussions and comments on 
the manuscript, and also for
bringing references \cite{miyashita}  and \cite{diptiman}   to notice.

\narrowtext

\begin{figure}

\caption{The local magnetisation shows an oscillatory behaviour about (a) 
a non-zero value for small  fluctuations  (b)   
zero  for large  fluctuations. The amplitude of the oscillations decreases
after reaching a maximum. The data are shown for a $N = 16$ system with 
$\lambda = 8$ and $h_0/J = 0.1$}    

\caption{Typical behaviour  of  the response to a spatially periodic 
 external field as a function of the quantum 
fluctuations for different wavelengths are shown. 
The values of $h_0/J$ are small 
($\sim 10^{-1}$) and different  
 for each of the curves and no systematics is expected. 
The maximum value of $g$ indicates the resonance.}

\caption{The $\Gamma_R/J$ lines for different wavelengths. The values of
$h_0/J$ where $\Gamma_R/J \rightarrow 0$ correspond to $h_0/J = 4\lambda$.
The solid line denotes the phase boundary $\Gamma_c$ for $\lambda = 2$}

\caption{The maximum value of the response is shown against the
amplitude of the field. It shows a linear behaviour upto a limiting value
of the field.}

\caption{The values of $\Gamma_R /J$ are shown against $\lambda$ keeping 
$h_0\lambda/J$ constant
in order to compare the results. The behaviour is non-monotonic for 
large $h_0 \lambda /J$.}
\end{figure}
\end{multicols}

\end{document}